\pdfoutput=1

\documentclass{emulateapj}
\usepackage{amsmath, amssymb}
\usepackage{color}
\usepackage[normalem]{ulem}
\usepackage{bm}

\slugcomment{Not to appear in Nonlearned J., 45.}

\shorttitle{Blocking metal accretion onto PopIII stars by stellar wind}
\shortauthors{S. J. Tanaka, G. Chiaki, N. Tominaga, \& H. Susa}

\begin{document}


\title{Blocking metal accretion onto population III stars by stellar wind}


\author{Shuta J. Tanaka\altaffilmark{1}}
\author{Gen Chiaki\altaffilmark{1}}
\author{Nozomu Tominaga\altaffilmark{1,2}}
\author{Hajime Susa\altaffilmark{1}}
\altaffiltext{1}{
Department of Physics, Faculty of Science and Engineering, Konan University, 8-9-1 Okamoto, Kobe, Hyogo 658-8501, Japan
}
\altaffiltext{2}{
Kavli Institute for the Physics and Mathematics of the Universe (WPI), The University of Tokyo, 5-1-5 Kashiwanoha, Kashiwa, Chiba 277-8583, Japan\altaffilmark{2}
}
\email{sjtanaka@center.konan-u.ac.jp}

\begin{abstract}
	Low-mass population III (PopIII) stars of $\lesssim 0.8 M_{\odot}$ could survive up until the present.
	Non-detection of low-mass PopIII stars in our Galaxy has already put a stringent constraint on the initial mass function (IMF) of PopIII stars, suggesting that PopIII stars have a top-heavy IMF.
	On the other hand, some claims that the lack of such stars stems from metal enrichment of their surface by accretion of heavy elements from interstellar medium (ISM).
	We investigate effects of the stellar wind on the metal accretion onto low-mass PopIII stars because accretion of the local ISM onto the Sun is prevented by the solar wind even for neutrals.
	The stellar wind and radiation of low-mass PopIII stars are modeled based on knowledge of nearby low-mass stellar systems including our Sun.
	We find that low-mass PopIII stars traveling across the Galaxy forms the stellar magnetosphere in most of their life.
	Once the magnetosphere is formed, most of neutral interstellar particles are photoionized before reaching to the stellar surface and are blown away by the wind.
	Especially, the accretion abundance of iron will be reduced by a factor of $< 10^{-12}$ compared with Bondi-Hoyle-Lyttleton accretion.
	The metal accretion can enhance iron abundance [Fe/H] only up to $\sim -14$.
	This demonstrates that low-mass PopIII stars remain pristine and will be found as metal free stars and that further searches for them are valuable to constrain the IMF of PopIII stars.
\end{abstract}

\keywords{early Universe --- first stars --- stars: low-mass --- stars: Population III --- stars: abundances --- stars: chemically peculiar}

\section{INTRODUCTION}\label{sec:intro}

Formation of first stars is one of the most important issue in the modern cosmology.
It has been revealed that most of them are as massive as $10-1000 M_{\odot}$, on the basis of $\Lambda$CDM cosmology \citep{Omukai&Nishi98,Abel+02,Bromm&Larson04,Yoshida+08,Stacy+11,Hosokawa+11,Stacy+12,Bromm13,Susa13,Susa+14,Hirano+14,Hirano+15}.
However, these stars are unlikely to be directly observed even with the next generation facilities.

On the other hand, low-mass first stars could also be formed via the fragmentation of the circumstellar disk around the primary proto-first-stars \citep{Clark+11a,Clark+11b,Greif+11,Greif+12}.
If the fragments could escape from the disk, they will survive to be low-mass stars although the final fate of them is still under debate \citep{Machida&Doi13,Stacy+16,Hirano&Bromm16}.
If these stars are less massive than $0.8 M_{\odot}$, they survive up until the present to be found as low-mass Population III (PopIII) stars in our Galactic halo.

The efforts to find the survived low-mass PopIII stars have been made continuously since the last century, but no such star has been found so far, although more than $10^5$ halo stars have already surveyed \citep[][and the references therein]{Frebel&Norris15}.
If this deficiency of low-mass PopIII stars is intrinsic, the present observations have already put a rather stringent constraint on the low-mass end of the initial mass function (IMF) of PopIII stars \citep{Hartwig+15,Ishiyama+16}, which proves very different nature of primordial star formation process from the present-day counterpart.

There is another interpretation of the lack of low-mass PopIII stars.
Their surface might be enriched by heavy elements in interstellar medium (ISM).
The pristine low-mass PopIII stars travel across the Galaxy after their formation and potentially accrete materials onto their surface \citep{Yoshii81, Iben83, Yoshii+95, Frebel+09}.
Since the convective layer of these stars in the main-sequence (MS) phase is shallow, the surface enrichment could be responsible for the observed element abundances even if the accreted masses of the metals are small.
Recent semi-analytical/numerical calculations suggest that the surface of low-mass PopIII stars are normally stained by metals up to the level of [Fe/H] $\sim -5$ \citep{Komiya+15,Shen+16}, and [Fe/H] $\sim -2$ for the most extreme cases \citep{Shen+16}.
These results suggest that the low-mass PopIII stars could be hidden in the observed metal poor stars.

However, these calculations assume Bondi-Hoyle-Lyttleton (BHL) accretion of the ISM onto the star \citep{Hoyle&Lyttleton39,Bondi52} and ignore the effects of the stellar magnetosphere \citep[c.f., however,][]{Vauclair&Charbonnel95,Johnson&Khochfar11}.
For example, our Sun has the magnetosphere called `heliosphere' extending out to $\gtrsim 120~{\rm AU} \approx 3 \times 10^4 R_{\odot}$ \citep[c.f.,][]{Burlaga+13, Gurnett+13, Krimigis+13, Stone+13, Burlaga&Ness16}.
Despite the collisionless nature of the solar wind plasma, there is a boundary called `heliopause' that separates the solar wind and the ISM plasmas electromagnetically \citep[c.f.,][]{Colburn&Sonett66}.
As shown in Figure \ref{fig:Sketch}, the ionized local ISM drapes around the heliopause and does not accrete onto the Sun, while the neutral components penetrate relatively freely into the heliosphere \citep[e.g.,][]{Patterson+63}.

Actually, neutral hydrogen, helium and even heavier elements originated from the local ISM have been observed around the Earth's orbit $r_{\rm E} = 1$ AU \citep[e.g.,][]{Bertaux&Blamont71, Weller&Meier74, Mobius+09}.
However, it is known that the abundance ratio at $r_{\rm E}$ is totally different from that of the original local ISM \citep[][]{Bochsler+12} because the solar radiation field and solar wind ionize the neutrals, and the efficiency of the ionization processes depends on elements \citep[c.f.,][]{Axford72}.
For example, neutral helium penetrates more deeply inside the heliosphere ($\sim 0.3$ AU) than neutral hydrogen ($\sim$ AU) because of the higher ionization potential of helium than that of hydrogen.

Once the `originally neutral' interstellar particles are ionized inside the heliosphere, their behavior is rather complicated.
However, these ions are immediately trapped by the magnetic field frozen into the solar wind, i.e., they are blown away from the Sun rather than accrete onto it \citep[c.f.,][]{Wu&Davidson72, Vasyliunas&Siscoe76, Isenberg86, Lee&Ip87}.
In the heliosphere, these ions have been directly observed as `pickup ions' \citep[e.g,][]{Mobius+85, Gloeckler+93} and as anomalous cosmic-rays \citep[e.g.,][]{Garcia-Munoz+73, Fisk+74}.

The existence of the heliosphere prevents the simple BHL accretion of the neutral ISM particles onto the Sun.
We extend this discussion to low-mass PopIII stars.
Accretion of heavy elements from the ISM onto PopIII stars is not so simple as in the previous studies.

In this paper, we examine the accretion rate of heavy elements onto low-mass PopIII stars, taking special care of the wind from the star and the ionization of the neutrals in the ISM.
In Section \ref{sec:MagnetosphereFormation}, we model the wind of low-mass PopIII stars and the surrounding ISM, and study the condition in order to form the magnetosphere around the stars.
In Section \ref{sec:Photoionization}, we investigate photoionization of the neutral ISM by the stellar radiation, and obtain the fraction which attains to the stellar surface.
Discussions for further studies are made in Section \ref{sec:Discussion} and we conclude this paper in Section \ref{sec:Conclusions}.
Throughout this paper, the steady and spherically symmetric wind zone is assumed, although we take into account the variations of stellar wind and the surrounding ISM parameters.

\section{Formation of Magnetosphere by Stellar Wind}\label{sec:MagnetosphereFormation}

%
\begin{figure}
\begin{center}
	\includegraphics[scale=0.3]{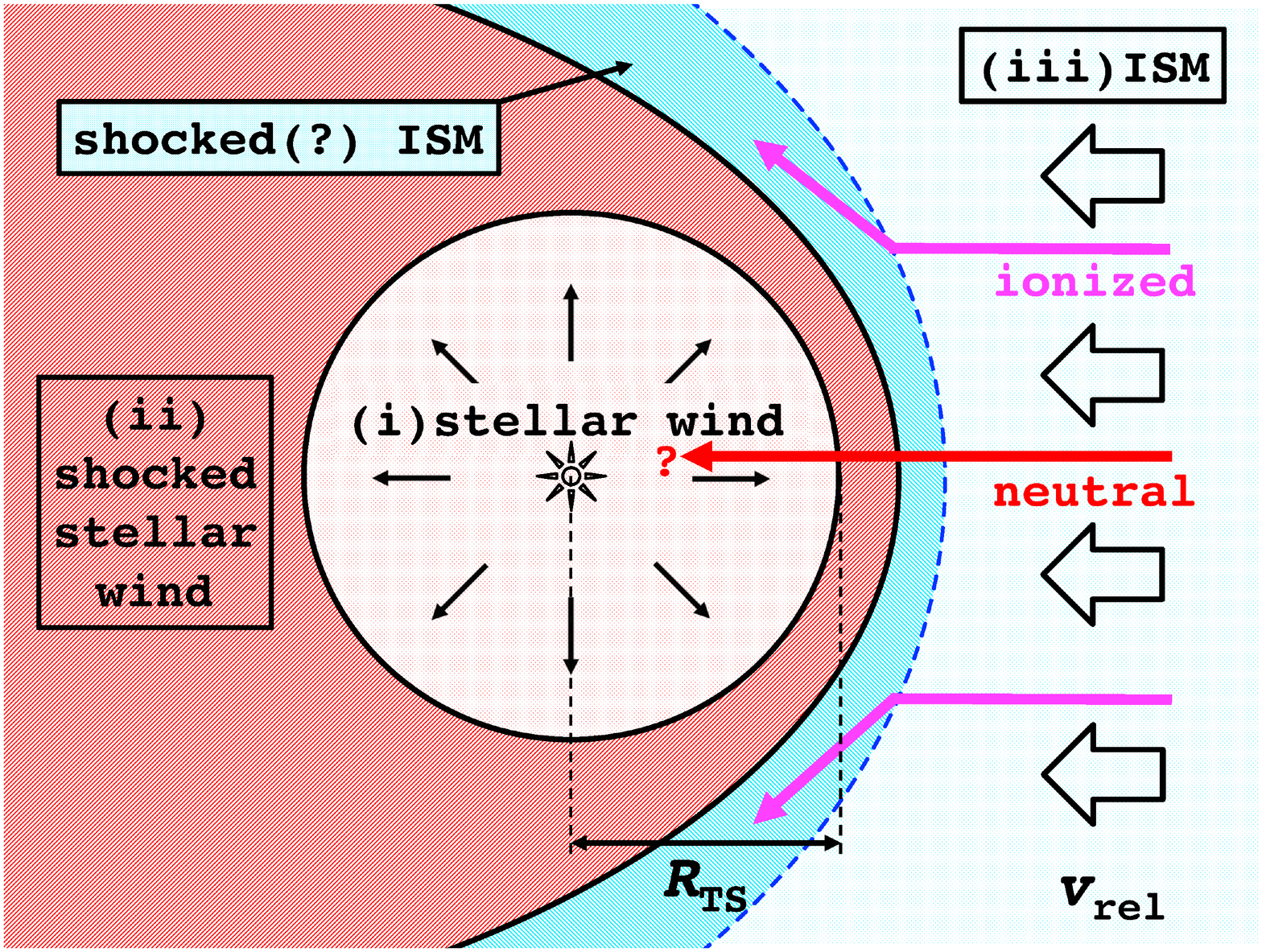}
\end{center}
\caption{
	Schematic picture of a stellar magnetosphere.
	We consider a star moving through the ISM of a density $n_{\rm ISM}$ with a relative velocity of $v_{\rm rel}$.
	The interaction between the stellar wind and the ISM forms at least three thermodynamically distinct regions (i) $-$ (iii).
	The region immediately around the star is the supersonic wind zone and is assumed to be almost spherical (region (i)).
	The wind is shocked and the hot subsonic wind zone is formed (region (ii)).
	These red region is made of plasma supplied from the central star, while the blue region is made of the ISM plasma (region (iii)).
	The contact surface between these different plasma is called heliopause in the case of the Sun.
	The ionized particles do not across the contact surface, while the neutral interstellar particles can penetrate into the magnetosphere.
}
\label{fig:Sketch}
\end{figure}

We consider a star moving through the ISM of a density $n_{\rm ISM}$ with a relative velocity of $v_{\rm rel}$.
The interaction between the stellar wind and the ISM forms at least three thermodynamically distinct regions (Figure \ref{fig:Sketch}): (i) the cold supersonic stellar wind, (ii) the hot subsonic stellar wind (shocked stellar wind), and (iii) the surrounding ISM.
The regions (i) and (ii) are separated by the termination shock, whose radius is $R_{\rm TS}$, and the regions (ii) and (iii) are separated by the contact discontinuity.
If $v_{\rm rel}$ exceeds the sound velocity of the ISM, the bow shock is formed (dashed line).
This hydrodynamical picture is mediated by the magnetic field and then the neutral components of the ISM can cross the boundaries.

Before discussing the fate of the neutral interstellar particles (NISPs) penetrating into the magnetosphere (the region (i)), we should study whether the magnetosphere is formed around the low-mass PopIII stars or not.
If the stellar wind is not strong enough, the ISM accretion flow shrinks the magnetosphere, reachs the stellar surface and enrichs it with heavy elements.
Based on the model described in Section \ref{sec:FirstModel}, we estimate critical ISM densities to shrink the magnetosphere against the stellar wind in Section \ref{sec:FirstResults}.

\subsection{Model}\label{sec:FirstModel}

\subsubsection{Low-mass PopIII Stars}\label{sec:StellarModel}

%
\begin{figure}
\begin{center}
	\includegraphics[scale=0.7]{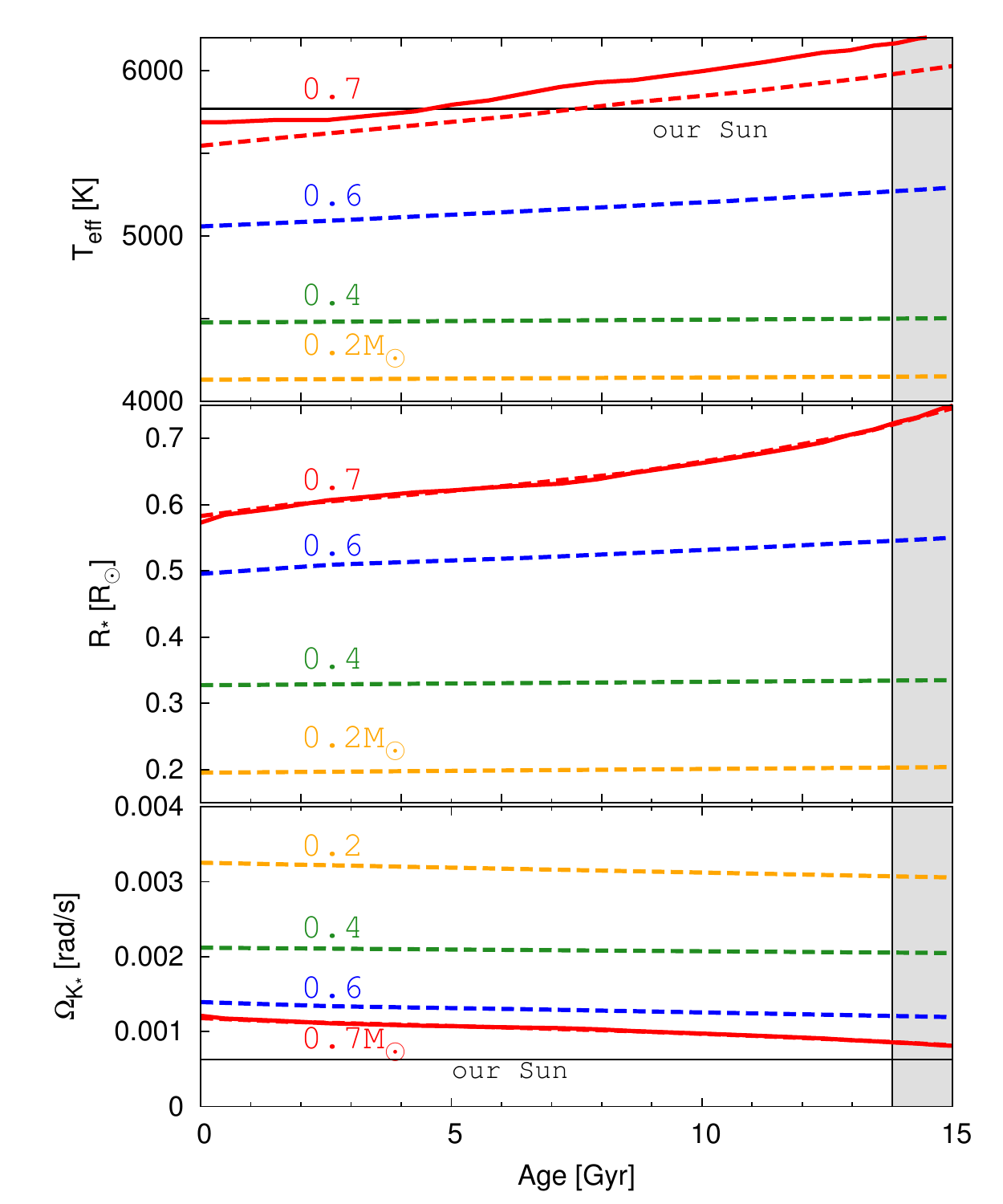}
\end{center}
\caption{
	Evolution of effective temperature $T_{\rm eff}$ (top panel), stellar radius $R_{\star}$ (middle panel) and the Kepler frequency at the stellar surface $\Omega_{\rm K \star}$ (bottom panel) for different masses, 0.7 (red), 0.6 (blue), 0.4 (green) and 0.2 $M_{\odot}$ (yellow).
	Red solid line is taken from \citet{Marigo+01} for 0.7 $M_{\odot}$ zero-metallicity star ($Z = 0$: solid lines), while dashed lines are $Z = 0.0004$ stars taken from \citet{Girardi+00}.
	However, the evolution of both $T_{\rm eff}$, $R_{\star}$ and $\Omega_{\rm K \star}$ is almost identical for $Z=0$ (red solid) and $0.0004$ (red dashed) for the case of 0.7 $M_{\odot}$.
	The values of our Sun are shown in grey lines for example.
	The shaded region is the age larger than the universe (13.8 Gyr) and the life-time of the star of $\ge 0.8 M_{\odot}$ is less than the age of the universe.
}
\label{fig:StellarModel}
\end{figure}
%

We study low-mass PopIII stars whose mass is $M_{\star} \le 0.7~M_{\odot}$ because they are in the MS phase during the cosmic history.
The stellar models are adopted from \citet{Marigo+01} for 0.7 $M_{\odot}$ zero-metallicity star ($Z = 0$), while, for the stars of smaller masses, we adopt $Z = 0.0004$ stars from \citet{Girardi+00}.
Figure \ref{fig:StellarModel} shows the key stellar parameters used in this paper; effective temperature $T_{\rm eff}$, stellar radius $R_{\star}$, and the Kepler frequency $\Omega_{\rm K \star} \equiv \sqrt{G M_{\star} / R^3_{\star}}$, where $G$ is the gravitational constant.
The two lines of 0.7 $M_{\odot}$ stars with different metallicities are fairly close to each other so that we also call the $Z = 0.0004$ stars as low-mass PopIII stars below.

Although the stellar parameters slightly change with time, we take the zero-age-main-sequence (ZAMS) values as the fiducial values.
0.8 $M_{\odot}$ stars from \citet{Marigo+01} are also marginally in the MS phase at an age of 13.8 Gyr but $R_{\star}$ at 13.8 Gyr is about an order of magnitude larger than that at ZAMS phase.
We do not show the result for 0.8 $M_{\odot}$ PopIII stars in this paper, but it is easy to extend the present study to the heavier mass of $> 0.8~M_{\odot}$ and also other stellar models.
Generally, the metal accretion onto stars are more difficult for more massive stars.

\subsubsection{Stellar Wind}\label{sec:StellarWind}


We apply the stellar wind model of low-mass MS stars to low-mass PopIII stars.
However, in contrast to the radiation driven wind from high-mass MS stars, the wind acceleration mechanism for low-mass MS stars is not fully understood, while some possible models are suggested \citep[e.g.,][]{Suzuki&Inutsuka05, Matsumoto&Suzuki14, vanBallegooijen&AsgariTarghi16}.
Although a direct detection of the stellar wind from low-mass MS stars has not been made, evolution of the stellar rotation period \citep[e.g.,][]{Barnes03, Bouvier+14, Matt+15} and also some indirect evidences of the stellar wind \citep[e.g.,][]{Gaidos+00, Kislyakova+14, Wood+14} indicate that they have a magnetized wind which extracts their angular momentum \citep[c.f.,][]{Weber&Davis67}.

In this paper, we impose that all the low-mass PopIII stars have a simple steady spherical stellar wind whose density and velocity profiles are
\begin{eqnarray}\label{eq:StellarWind}
	n_{\rm sw}(r)
	=
	n_{\rm sw \star}
	\left( \frac{r}{R_{\star}} \right)^{-2},~~
	v_{\rm sw}(r)
	=
	v_{\rm sw \star},
\end{eqnarray}
respectively \citep[e.g.,][]{Talbot&Newman77}.
Equation (\ref{eq:StellarWind}) approximately describes the thermal driven supersonic outflow beyond the sonic point \citep[][]{Parker58}.
In the case of the solar wind, the normalizations $(n_{\rm sw \star}, v_{\rm sw \star})$ of Equation (\ref{eq:StellarWind}) are obtained from the observed values at the Earth's orbit.
Although, in addition to the time-variabilities, the significant heliolatitude structure of the solar wind is known \citep[e.g.,][]{Phillips+95, Sokol+13}, we adopt the value of the slow wind, $n_{\rm sw}(r_{\rm E}) = 5~{\rm cm^{-3}}$ and $v_{\rm sw}(r_{\rm E}) = 400~{\rm km~s^{-1}}$ as the fiducial values \citep[e.g.,][]{King&Papitashvili05, Bzowski+13b}.

For the value of $v_{\rm sw \star}$ of low-mass PopIII stars, we adopt the solar value irrespective of the stellar mass, i.e., $v_{\rm sw \star} = 400~{\rm km~s^{-1}}$.
The wind is supersonic and its velocity should be higher than the sound velocity of their coronas $\sim$ 300 km s$^{-1}$ for the temperature of $\sim$ keV \citep[e.g.,][]{Gudel04, DeWarf+10}.
The wind velocity would be comparable with the escape velocity from stars $v_{\rm esc \star}$, or precisely, that from the coronal radius $\sim$ a few $R_{\star}$.
The values of $v_{\rm esc \star} \equiv \sqrt{2} R_{\star} \Omega_{\rm K \star}$ calculated from the stellar model of Figure \ref{fig:StellarModel} are similar among different masses and fall within the range of $600~{\rm km~s^{-1}} < v_{\rm esc \star} < 700~{\rm km~s^{-1}}$.
The solar value is $v_{\rm esc \odot} \approx 620~{\rm km~s^{-1}}$.

The value of $n_{\rm sw \star}$ is found from the observations of the mass loss rate $\dot{M}_{\star}$ combined with $v_{\rm sw \star} = 400~{\rm km~s^{-1}}$.
However, $\dot{M}_{\star}$ for low-mass MS stars is difficult to observe because it is low.
For example, the mass loss rate of the Sun is only $\dot{M}_{\odot} = 4 \pi r^2_{\rm E} m_{\rm p} n_{\rm sw \odot}(r_{\rm E}) v_{\rm sw \odot} \sim 10^{-14}~M_{\odot}~{\rm yr^{-1}}$, where the main component of the stellar wind is set to hydrogen throughout this paper.
The mass loss rate of $\dot{M}_{\star} = \dot{M}_{\odot}$ is taken as the fiducial value, irrespective of the mass of low-mass PopIII stars.
On the other hand, the mass loss rate of low-mass MS stars does not only depend on their mass but also depends on their rotation, age and magnetic field.
For example, a series of observations \citep[][]{Wood+01,Wood+02,Wood+05b,Wood+14} indicates $0.1 \dot{M}_{\odot} < \dot{M}_{\star} < 100 \dot{M}_{\odot}$ for some low-mass MS stars.
We take into account an order of magnitude variation of $n_{\rm sw \star}$ around the fiducial value as an uncertainty of the mass loss rate.

\subsubsection{Surrounding Interstellar Medium}\label{sec:ISM}

The termination shock of our solar system is located about 100 AU away from the Sun \citep[][]{Burlaga+05, Decker+05, Gurnett&Kurth05, Stone+05}, which is consistent with the estimate from the pressure balance condition between the solar wind ram pressure $\approx m_{\rm p} n_{\rm sw}(r) v^2_{\rm sw}$ and the interstellar magnetic pressure with $B_{\rm ISM} \approx 5~\mu{\rm G}$ \citep[e.g.,][]{Axford+63, Thomas78, Rand&Kulkarni89}.
Although the total pressure of the local ISM around the Sun is dominated by the magnetic pressure, the situation depends on stellar environments.
Especially, the ram pressure of the local ISM can be much higher than the magnetic pressure for large $n_{\rm ISM}$ and $v_{\rm rel}$, and has potential to be large enough to shrink the termination shock radius $R_{\rm TS}$ down to the stellar radius $R_{\star}$.

We assume BHL accretion of the surrounding ISM \citep{Hoyle&Lyttleton39,Bondi52,Shima+85}.
Introducing the critical impact parameter for BHL accretion $\xi_{\rm BHL} = 2 G M / v^2_{\rm rel} = R_{\star} (v_{\rm esc \star} / v_{\rm rel})^2$, the BHL accretion rate is written as $\dot{M}_{\rm BHL} = \pi \xi^2_{\rm BHL} m_{\rm p} n_{\rm ISM} v_{\rm rel}$, where the main component of the ISM is also set to hydrogen throughout this paper.
To be precise, we should write $\sqrt{v^2_{\rm rel} + c^2_{\rm s}}$ instead of $v_{\rm rel}$ in the expressions of $\xi_{\rm BHL}$ and $\dot{M}_{\rm BHL}$, where $c_{\rm s}$ is the sound velocity of the ISM.
For $v_{\rm rel} \rightarrow 0$, $\dot{M}_{\rm BHL}$ approaches spherical Bondi accretion.
The density and velocity of the accretion flow are \citep[c.f.,][]{Bisnovatyi-Kogan+79, Edgar04}
\begin{eqnarray}\label{eq:HoyleLyttletonTrajectory}
	n_{\rm HL}(r, \theta, \xi)
	=
	\frac{n_{\rm ISM} \xi^2}{r \sin \theta (2 \xi - r \sin \theta)},~~ \\
	v_{{\rm HL},r} (r, \theta, \xi)
	=
	- \sqrt{ v^2_{\rm rel} \left(1 - \frac{\xi^2}{r^2} \right) + v^2_{\rm esc \star} \frac{R_{\star}}{r}},
\end{eqnarray}
where $\xi$ is an impact parameter, and the boundary conditions $n_{\rm ISM}$ and $-v_{\rm rel}$ are given at $r \rightarrow \infty$ and $\theta \rightarrow \pi$.

For example, the local ISM of our Sun is $n_{\rm ISM \odot} \sim 0.2~{\rm cm^{-3}}$ and $v_{\rm rel \odot} \sim 20~{\rm km~s^{-1}}$ at present \citep[c.f.,][]{Heerikhuisen+14} and \citet{Shen+16} found, in their numerical simulations, that the environment of low-mass PopIII stars is $n_{\rm ISM} \lesssim 10^2~{\rm cm^{-3}}$ and $v_{\rm rel} \lesssim 200~{\rm km~s^{-1}}$ in their entire life.
As fiducial values, we adopt $n_{\rm ISM} = 1~{\rm cm^{-3}}$ and $v_{\rm rel} = 200~{\rm km~s^{-1}}$, where the former corresponds to the average ISM density \citep[e.g.,][]{Draine11} and the latter corresponds to the typical velocity dispersion of the Galactic halo (metal poor) stars \citep[e.g.,][]{Chiba&Beers00}, respectively.
Note that the sound velocity of the ISM $c_{\rm s} \approx 20~{\rm km~s^{-1}}$ is much smaller than $v_{\rm rel}$ of halo stars \citep[e.g.,][]{Draine11}.

\subsection{Results}\label{sec:FirstResults}


We presume the hydrodynamic interaction of the surrounding ISM and the stellar wind in this section, i.e., we regard the surrounding ISM as fully ionized in this section.
The radius of the termination shock $R_{\rm TS}$ is obtained by equating the ram pressures of the stellar wind and the accretion flow.
Considering the accretion flow almost along with the axis ($\sin \theta = \xi / r \ll 1$), $R_{\rm TS}$ would be calculated from the condition
\begin{eqnarray}\label{eq:TSRadiusCondition}
	n_{\rm sw \star} v^2_{\rm sw \star} \left( \frac{R_{\star}}{R_{\rm TS}} \right)^2
	& \approx &
	n_{\rm ISM} \left( v^2_{\rm rel} + v^2_{\rm esc \star} \frac{R_{\star}}{R_{\rm TS}} \right).
\end{eqnarray}
%

Following \citet{Talbot&Newman77}, we require the condition $R_{\rm TS} > \xi_{\rm BHL}$ for the formation of the magnetosphere (see also Section \ref{sec:Discussion} for this condition).
The critical density for magnetosphere formation becomes
\begin{eqnarray}\label{eq:CriticalDensity}
	n_{\rm crit}
	& \equiv &
	\frac{n_{\rm sw \star}}{2} \frac{v^2_{\rm sw \star} v^2_{\rm rel}}{v^4_{\rm esc \star}} \nonumber \\
	& \approx &
	10^4~{\rm cm^{-3}}
	\left( \frac{n_{\rm sw \star}}{7.0 \times 10^5~{\rm cm^{-3}}} \right)
	\left( \frac{v_{\rm sw \star}}{400          ~{\rm km~s^{-1}}} \right)^{ 2} \nonumber \\
	& &
	\left( \frac{v_{\rm rel     }}{200          ~{\rm km~s^{-1}}} \right)^{ 2}
	\left( \frac{v_{\rm esc\star}}{680          ~{\rm km~s^{-1}}} \right)^{-4}
\end{eqnarray}
%
where we adopt the parameters of a 0.7 $M_{\odot}$ PopIII star.
When $n_{\rm ISM} < n_{\rm crit}$, there is the magnetosphere extending $R_{\rm TS} > \xi_{\rm BHL} (> R_{\star})$ around the star.

Allowing an order of magnitude variation of $n_{\rm sw \star}$ ($\dot{M}_{\star}$), the critical ISM density is $10^3~{\rm cm^{-3}} \lesssim n_{\rm crit} \lesssim 10^5~{\rm cm^{-3}}$.
Although $n_{\rm crit}$ decreases with $v_{\rm rel}$, $v_{\rm rel} \ll 200~{\rm km~s^{-1}}$ is expected only in the early universe (see discussion in Section \ref{sec:Overdence}).
Note that Equation (\ref{eq:CriticalDensity}) is essentially the same as Equation (26) of \citet{Talbot&Newman77}, where they ignored the second term of the left-hand side of Equation (\ref{eq:TSRadiusCondition}).

According to the results of \citet{Shen+16}, the density of the ISM around low-mass PopIII stars is $n_{\rm ISM} \lesssim 10^2~{\rm cm^{-3}} \ll n_{\rm crit}$, i.e., they always have the magnetosphere extending to $R_{\rm TS} / R_{\star} > 10^2$ from Equation (\ref{eq:TSRadiusCondition}).
Because it is difficult to resolve the region $n_{\rm ISM} > n_{\rm crit}$ by the current cosmological simulations, the metal enrichment by the `ionized' ISM is hardly expected for the simulation done by \citet{Shen+16}.
However, we know that the ISM does have such dense region like molecular clouds, we discuss the accretion from the region $n_{\rm ISM} > n_{\rm crit}$ below.

\subsection{Discussion}\label{sec:FirstDiscussion}

\subsubsection{Accretion from Overdense Regions}\label{sec:Overdence}

Introducing the probability distribution function (PDF) of the ISM density $P(n,t)$ and the metallicity distribution $Z(n,t)$, we write the amount of the accretion metal mass
\begin{eqnarray}\label{eq:AccretionMass}
	M_{Z,{\rm acc}}
	& = &
	\int dt \int^{\infty}_{n_{\rm crit}(t)} d n P(n,t) Z(n,t) \dot{M}_{\rm BHL}(n, t).
\end{eqnarray}
In general, the mass of the stellar convective layer $M_{\rm conv} = f_{\rm conv} M_{\star}$ is much larger than the total accretion mass, where $f_{\rm conf} \sim 10^{-3}$ for $M_{\star} = 0.7~M_{\odot}$ \citep[][]{Yoshii81}.
The present metal abundance acquired by BHL accretion would be estimated as [Z/H]$_{\rm acc} \approx \log (M_{Z,{\rm acc}} / M_{\rm conv})$.
\citet{Shen+16} evaluated Equation (\ref{eq:AccretionMass}) numerically but they set $n_{\rm crit} = 0$, i.e., the case of no stellar wind.

The evaluation of Equation (\ref{eq:AccretionMass}) is fairly difficult in both numerical and analytical ways.
The density PDF of the ISM stems from its multi-phase nature, and high-density clouds are formed as a result of highly nonlinear physics including shocks, turbulences, thermal instability, self-gravity, and also magnetic field \citep[c.f.,][]{Koyama&Inutsuka00, Koyama&Inutsuka02, Inoue&Inutsuka08, Inoue&Inutsuka09, Inoue&Inutsuka12, Hennebelle+08, Banerjee+09}.
The thermal instability is important for the initial fragmentation of the diffuse ISM and depends on metallicity of the ISM \citep[][]{Inoue&Omukai15}.
For metallicity distribution, $Z(n,t)$ is increasing function of $t$ and would be decrease function of $n$ in the early universe \citep[e.g.,][]{Ritter+12, Smith+15, Chen+16}

Despite the above difficulties, \citet{Johnson&Khochfar11} have studied metal enrichment of low-mass PopIII star $M_{Z,{\rm acc}}$ taking into account a finite $n_{\rm crit}$ obtained by \citet{Talbot&Newman77}.
They considered high redshift ($z \sim$ 10) universe because the slower relative velocity $v_{\rm rel}$, i.e., larger $\dot{M}_{\rm BHL}$ and smaller $n_{\rm crit}$, than the present is expected.
They found that an encounter probability of a star with a dense cloud is less than 0.1 and, in addition, metal enrichment by one interaction of a star with a dense cloud is only [Fe/H]$_{\rm JK11} \sim -6$ even adopting an order of magnitude larger metallicity $Z \sim 10^{-2} Z_{\odot}$ of the surrounding ISM from \citet{Shen+16} than their Equation (12).
[Fe/H]$_{\rm JK11}$ is marginal or insufficient to explain the observed lowest [Fe/H] $\sim -5$ \citep[][]{Christlieb+04, Aoki+06, Caffau+11, Frebel+15}.


\subsubsection{Roles of Neutrals}\label{sec:RolesOfNeutrals}

Equation (\ref{eq:TSRadiusCondition}) is applicable for the ionized component of the ISM.
The average ionization fraction of the ISM is only $\sim$ 0.1 \citep[e.g.,][]{He+13} and the ionization fraction is much lower than the average at dense regions like molecular clouds \citep[e.g.,][]{Draine11}.
The simple hydrodynamical interaction between the mostly neutral ISM and the fully ionized stellar wind would be inappropriate treatment because they are collisionless as depicted in Figure \ref{fig:Sketch} (see also discussion in Section \ref{sec:Discussion}).
Accretion of NISPs penetrating into the magnetosphere will be discussed in Section \ref{sec:Photoionization}.

\section{Photoionization of NISPs Penetrating into Magnetosphere}\label{sec:Photoionization}

In the previous section, we focus on the interaction between the stellar wind and the ionized components of the ISM in terms of the formation of the magnetosphere.
As is already shown in Figure \ref{fig:Sketch}, even if the termination shock of the stellar wind is formed, i.e., $n_{\rm ISM} < n_{\rm crit}$, NISPs go through the magnetosphere and has the accretion trajectory described in Equation (\ref{eq:HoyleLyttletonTrajectory}).
However, by approaching the stellar surface, they are suffered from the ionization processes, especially photoionization.
Before studying photoionization of NISPs penetrating into the magnetosphere, we briefly mention the fate of NISPs ionized inside the magnetosphere.

\subsection{Model}\label{sec:SecondModel}

\subsubsection{Trapping Ionized NISPs into Stellar Wind}\label{sec:Trapping}

The magnetic field of the stellar wind $B_{\rm sw}$ plays a crucial role in trapping of NISPs into the stellar wind.
Once NISPs are ionized inside the magnetosphere, they are accelerated by the motional electric field of the stellar wind ${\bm E} = - ({\bm v}_{\rm sw} - {\bm v}_{\rm HL})/ c \times {\bm B}_{\rm sw}$ and gyrate around the magnetic field in the frame of the stellar wind, where $c$ is the speed of light.
The ratio of the gyro-radius $r_{\rm g}$ of a singly ionized iron (FeII) to the stellar radius is
%
\begin{eqnarray}\label{eq:GyroRatio}
	\frac{r_{\rm g,FeII}(R_{\star})}{R_{\star}}
	& \approx &
	10^{-5}
	\left( \frac{B_{\rm sw     \star}}{1~{\rm G}}                \right)^{-1}
	\left( \frac{\Omega_{\rm K \star}}{10^{-3}~{\rm rad~s^{-1}}} \right),
\end{eqnarray}
where we assume $v_{\rm esc \star} \gg v_{\rm rel}$ in order to evaluate the momentum of an FeII ion.
Note that an FeII ion has the largest mass-to-charge ratio, i.e., the largest gyro-radius, among the ions of our interest.
The magnetic field strength at the solar surface is $B_{\rm sw \odot} \sim$ 1 G on average and that at $r_{\rm E}$ is $\sim 10~{\rm \mu G}$.
Although the ratio $r_{\rm g,FeII}(r) / r$ would increase with $r$, only $r_{\rm g,FeII}(R_{\star}) / R_{\star} \ll 1$ is enough to trap ionized NISPs before reaching the stellar surface.

Equation (\ref{eq:GyroRatio}) indicates that a few orders of magnitude smaller surface magnetic field than that of the Sun can trap the photoionized NISPs into the stellar wind.
Below, we presume that low-mass PopIII stars have the surface magnetic field which satisfies $r_{\rm g,FeII}(R_{\star}) / R_{\star} \ll 1$.
In addition, the stellar wind has the power large enough to blown all the ionized NISPs away because we consider a situation $n_{\rm ISM} < n_{\rm crit}$.
What we need to study is only whether NISPs are ionized before reaching the stellar surface or not.


\subsubsection{Ionization Processes}\label{sec:IonizationProcesses}

NISPs are suffered from photoionization by stellar radiation, from ionization by charge exchange with stellar wind ions (mainly proton), and from electron impact ionization by stellar wind electrons \citep[c.f.,][]{Zank99}.
Photoionization is the dominant ionization process for heavy elements in the heliosphere \citep[c.f.,][]{Cummings+02}, and we study this process in the magnetosphere of low-mass PopIII stars.
The latter two ionization processes are less investigated \citep[c.f.,][]{Bzowski+13a} and omit them in this paper for simplicity.
We will give the ionization rates by charge exchange for some elements in Section \ref{sec:ChargeExchange} just for reference.


The local photoionization rate for the $i$th element is given by
\begin{eqnarray}\label{eq:PhotoionizationRateDefinition}
	\beta_{{\rm ph},i}(r)
	& = &
	\int d \lambda \dot{N}_{\lambda}(r) \sigma_{{\rm ph},i}(\lambda),
\end{eqnarray}
where $\dot{N}_{\lambda}(r)$ is the specific photon number flux density and $\sigma_{{\rm ph},i}(\lambda)$ is the photoionization cross section for the neutral $i$th element.
Here, we adopt the fitting formula of $\sigma_{{\rm ph},i}(\lambda)$ from \citet{Verner+96}.

Assuming that the entire stellar surface has a uniform and isotropic specific intensity $J_{\lambda}$ \citep[e.g.,][]{Rybicki&Lightman79}, we obtain
\begin{eqnarray}\label{eq:PhotonNumberFlux}
	\dot{N}_{\lambda}(r)
	& = &
	\pi \left( \frac{R_{\star}}{r} \right)^2
	\frac{J_{\lambda}}{h \nu},
\end{eqnarray}
where $\nu = c / \lambda$, $h$ is the Planck constant.
We obtain the inverse square $r$-dependence of $\beta_{{\rm ph},i}(r)$ from Equation (\ref{eq:PhotonNumberFlux}) as
\begin{eqnarray}
	\beta_{{\rm ph},i}(r)
	& = &
	\beta_{{\rm ph}\star,i}
	\left( \frac{R_{\star}}{r} \right)^2,~ \label{eq:PhotoionizationRate} \\
	\beta_{{\rm ph}\star,i}
	& \equiv &
	\int d \lambda \frac{\pi J_{\lambda}}{h \nu} \sigma_{{\rm ph},i}(\lambda). \label{eq:PhotoionizationRateAtSurface}
\end{eqnarray}
No depletion of photons by phoionization is assumed for this inverse square $r$-dependence.
The mean free path for the extreme ultraviolet (EUV) photons are more than 100 AU for the neutral density of a few cm$^{-3}$ because $\sigma_{{\rm ph},i} \lesssim 10^{-17}~{\rm cm^2}$ (c.f., the left panel of Figure \ref{fig:SolarSpectrum}).

\subsubsection{Stellar Radiation}\label{sec:StellarRadiation}

%
\begin{figure*}
\begin{center}
	\includegraphics[scale=0.7]{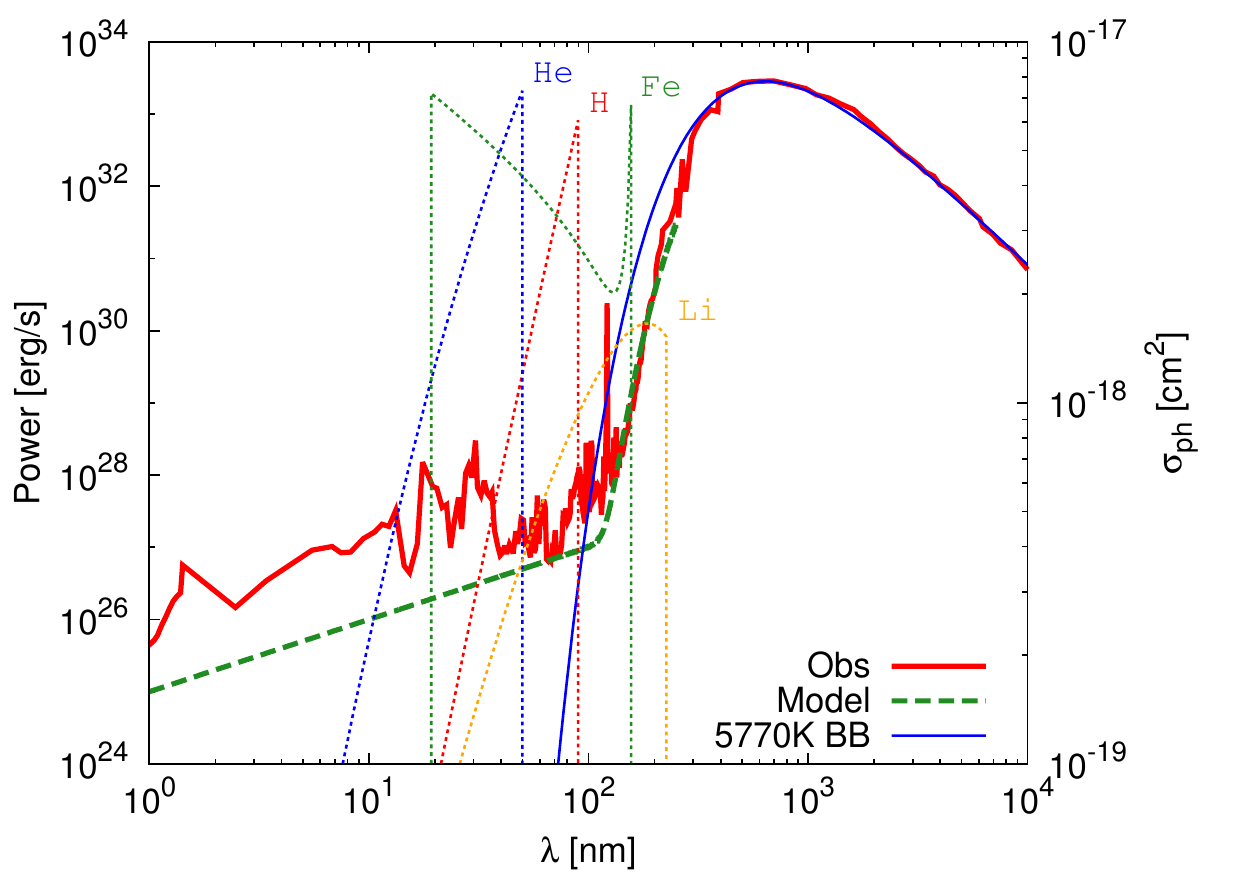}
	\includegraphics[scale=0.7]{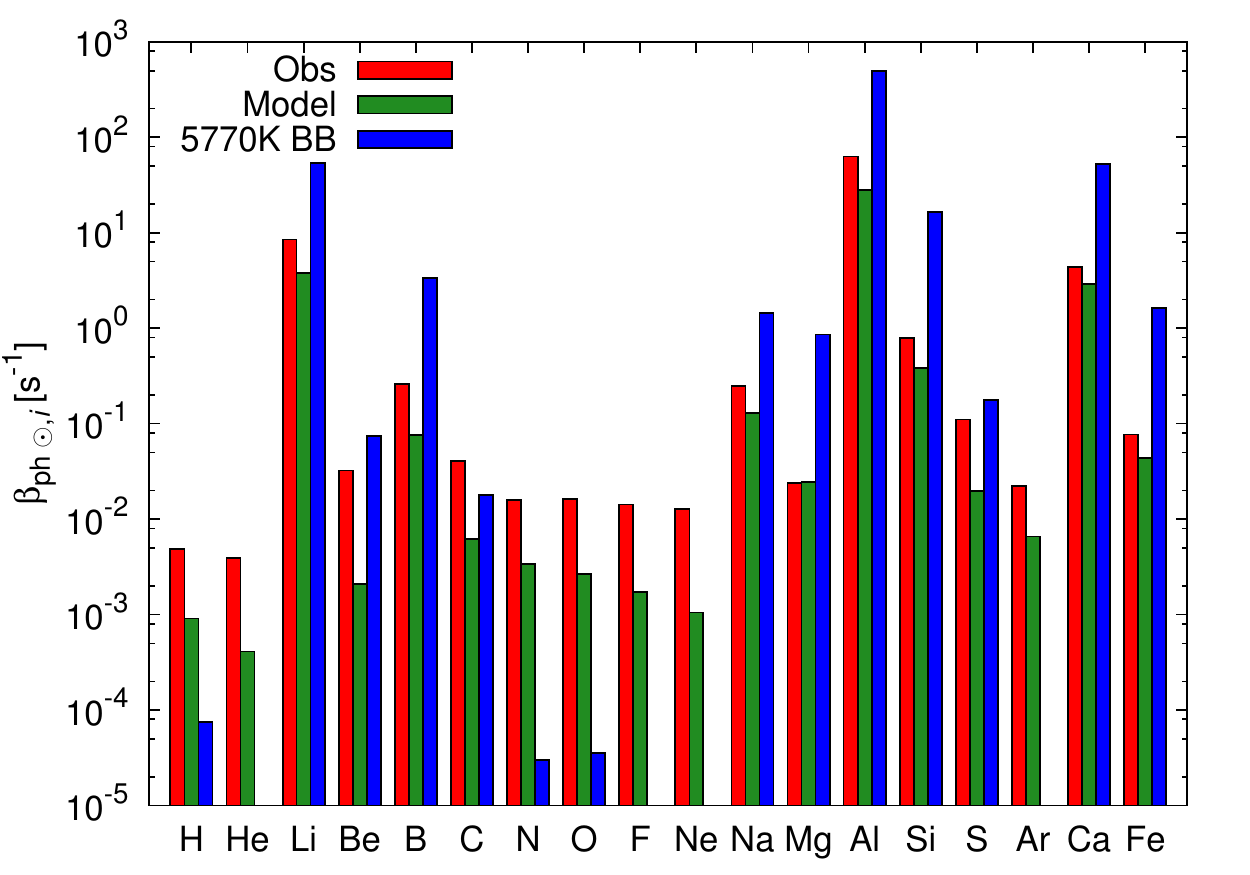}
\end{center}
\caption{
	(Left):
	Observational data of the solar spectrum (red line) is taken from \citet{Lean91, Woods+05, Caspi+15}.
	The photoionization cross sections for hydrogen, helium, lithium and iron are overplotted in dotted lines \citep[][]{Verner+96}.
	The spectrum responsible for photoionization does not agree with the 5770K blackbody (blue line: $\pi (r_{\odot} / r_{\rm E})^2 B_{\lambda}(5770K)$).
	Green line shows the model spectrum for UV radiation from the Sun (Equation (\ref{eq:IntensityOfPopIII}) with $T_{\rm eff} =$ 4721 K and $L_{\rm EUV,\lambda} = 10^{25}~{\rm erg~s^{-1}~nm^{-1}}$).
	(Right):
	The photoionization rates $\beta_{{\rm ph}\odot, i}$ (Equation (\ref{eq:PhotoionizationRateDefinition})) for all the elements that the photoionization cross sections are given by \citet{Verner+96}.
	Three different rates for each element correspond to the three difference of the radiation spectra in the left panel.
}
\label{fig:SolarSpectrum}
\end{figure*}

The stellar spectrum $J_{\lambda}$ is important for the estimate of $\beta_{{\rm ph}\star,i}$, since the first ionization potentials of all the elements are in EUV wavelength, i.e., the Wien regime of the blackbody spectrum for low-mass PopIII stars (see the top panel of Figure \ref{fig:StellarModel}).
Here, we model the spectrum of low-mass PopIII stars based on the spectra of the Sun and low-mass MS stars.

In the left panel of Figure \ref{fig:SolarSpectrum}, we show the spectral energy distribution (SED) of the Sun $\lambda L_{\lambda,\odot} \equiv 4 \pi r^2_{\rm E} \lambda F_{\lambda}(r_{\rm E})$.
$F_{\lambda}(r_{\rm E}) = h \nu \dot{N}_{\lambda}(r_{\rm E})$ is the observed specific solar flux density at the Earth's orbit \citep[][]{Lean91, Woods+05, Caspi+15}.
In order to illustrate the relevant wavelengths for photoionization, the photoionization cross sections for hydrogen, helium, lithium, and iron are overplotted with dotted lines \citep[][]{Verner+96}.
While the spectrum is very well approximated by the 5770 K blackbody $\gtrsim$ 400 nm, there are heavily absorbed features and many emission lines of the heavy elements at 100 nm $\lesssim \lambda \lesssim$ 400 nm \citep[e.g.,][]{Lean00,Curdt+01,Frohlich&Lean04}.
On the other hand, the emission below 100 nm is far dominated over the 5770 K blackbody by the power-law-like component originating from the chromosphere, transition region and corona.
The spectrum between 1 $-$ 100 nm seems to be approximated by the flat spectrum \citep[e.g., Figure 1 of][]{Ermolli+13}.

Before modeling SED of low-mass PopIII stars, in order to study effects of these complex spectral features at $\lesssim$ 400 nm, we calculate the photoionization rate of our Sun for the following three cases of the radiation spectra: (1) the observed spectrum (red in both left and right panels of Figure \ref{fig:SolarSpectrum}), (2) the model spectrum (green) and (3) the 5770 K blackbody (blue).
The radiation at $\gtrsim$ 250 nm does not contribute photoionization for all the elements listed in the right panel of Figure \ref{fig:SolarSpectrum}.
For the model spectrum (green), we fit the absorbed spectral feature at 130 nm $\le \lambda \le$ 250 nm with the Planck spectrum $\pi (r_{\odot} / r_{\rm E})^2 B_{\lambda}(\hat{T})$ and obtain $\hat{T} = 4721 \pm 15~{\rm K}$.
At $\lesssim$ 130 nm, we adopt the flat spectrum $L_{\lambda} = L_{\rm EUV, \lambda}$, which we call the EUV component hereafter.
We set the normalization as $\lambda L_{\rm EUV, \lambda} = 10^{27}~{\rm erg~s^{-1}}$ at $\lambda =$ 100 nm (see discussion below).

The right panel of Figure \ref{fig:SolarSpectrum} shows the resultant $\beta_{{\rm ph} \odot,i}$ for the three different radiation spectra.
The differences of $\beta_{{\rm ph} \odot,i}$ between the observed (red) and the model (green) spectra is less than an order of magnitude and we find that this crude model for the EUV component is sufficient for order-of-magnitude estimates of $\beta_{{\rm ph} \star, i}$.
The photoionization rate calculated by the 5770 K blackbody spectrum (blue) is far different from the observation (red), especially for the elements with the high ionization potential, such as noble gases.
On the other hand, for the 5770 K blackbody spectrum, some elements including lithium, aluminum, calcium, and iron have higher photoionization rates than for the case of the observed spectrum.

We assume that low-mass PopIII stars also have the EUV component in their SED.
The origin of the low-mass stellar wind would be closely related with the stellar X-ray emission through the coronal heating and the turbulent activity of the stellar convective layer \citep[e.g.,][]{Cranmer&Saar11, Suzuki+13}.
In the case of the Sun, the EUV component seems to have a peak at $\sim$ 100 nm and its luminosity is almost the same as the power of the solar wind $L_{\rm sw \odot} \approx \dot{M}_{\odot} v^2_{\rm sw \odot} \sim 10^{27}~{\rm erg~s^{-1}}$.
Interestingly, even the $\alpha$ Centauri system, which is the closest star system, harbors a solar-type star of the similar both X-ray luminosity \citep[][]{DeWarf+10} and $\dot{M}_{\star}$ \citep[][]{Wood+01} with those of the Sun.
In addition, the X-ray luminosity of young solar-type stars have been observed and some have three orders of magnitude higher X-ray luminosity than that of the Sun \citep[c.f.,][]{Gudel+97, Gudel04}.
The analyses by \citet{Linsky+13, Linsky+14} indicate that the EUV luminosity correlates positively with the X-ray luminosity for low-mass (F5 $-$ K5) MS stars.
Note that \citet{Linsky+14} found the positive correlation of the EUV luminosity with the Ly$\alpha$ luminosity for some low-mass MS stars and the Sun has the smallest Ly$\alpha$ luminosity among them.
Considering these facts, it is reasonable to assume that low-mass PopIII stars also have the EUV component of similar with or even higher flux level than the Sun.

In this paper, we assume that the specific intensity of low-mass PopIII stars is
\begin{eqnarray}
	\pi J_{\lambda}
	& = &
	\pi B_{\lambda}(T_{\rm eff})
	+
	F_{\rm EUV, \lambda},~ \label{eq:IntensityOfPopIII} \\
	F_{\rm EUV, \lambda}
	& \equiv &
	\frac{L_{\rm EUV, \lambda}}{4 \pi R^2_{\star}}. \label{eq:EUVFlux}
\end{eqnarray}
For low-mass PopIII stars, we expect that the Wien regime of their blackbody spectrum has no absorbed features by metals because their atmosphere is also metal free.
The fiducial value of the EUV component is set to the same level as the Sun $\lambda L_{\rm EUV, \lambda} = 10^{27} (\lambda / 100~{\rm nm}) ~{\rm erg~s^{-1}}$ at $\lambda =$ 100 nm, or $L_{\rm EUV, \lambda} = 10^{25}~{\rm erg~s^{-1}~nm^{-1}}$ independent from $\lambda$ (flat spectrum).
Within the assumption of the flat spectrum of the EUV component, the contribution of the EUV component to $\beta_{{\rm ph}\star, i}$ is exactly proportional to the value of $F_{\rm EUV, \lambda}$.
We remark that our model can be applicable to different stellar models ($T_{\rm eff},~R_{\star}$) and values of $L_{\rm EUV, \lambda}$.

\subsubsection{Rate Equation}\label{sec:RateEquation}

Solar-type stars are capable to form the ionized environment although the region is inside their magnetosphere being different from that around massive stars \citep[][]{Stromgren39}.
We consider the ionization of NISPs drifting into the magnetosphere of low-mass PopIII stars.
For simplicity, we study NISPs whose trajectory is along the line of the stellar motion, i.e., zero impact parameter, from the upwind direction.
This specific trajectory gives a lower limit on the ionization fraction of NISPs inside the magnetosphere because this minimizes the duration of the interaction between NISPs and ionizing photons.
The orbit is a simple one-dimensional motion in the stellar gravitational field
\begin{eqnarray}\label{eq:Trajectory}
	v(r)
	& = &
	- \sqrt{ v^2_{\infty} + v^2_{\rm esc \star} \frac{R_{\star}}{r}}.
\end{eqnarray}
We take the boundary condition of $v_{\infty} = v_{\rm rel}$.
In Equation (\ref{eq:Trajectory}), we do not include the radiation pressure force because it is negligible at least for some heavy elements \citep[c.f.,][]{Brasken&Kyrola98, Sokol+15}.

The ionization state is calculated from the rate equation.
We use the steady state approximation for simplicity.
Note that, for the case of our Sun, effects of the solar variabilities on the ionization rate are less than a factor of two \citep[c.f., Figure 3 of][]{Bzowski+13a}.
The density of the $i$th element of NISPs $n_i(r)$ is obtained from
\begin{eqnarray}\label{eq:IonizationEquation}
	v(r) \frac{d n_i(r)}{d r}
	& = &
	- \beta_{{\rm ph},i}(r) n_i(r),
\end{eqnarray}
where the recombination processes are omitted because the radiative and dielectric recombination rates are much lower than the ionization rate \citep[c.f.,][]{Grzedzielski+10}.
We can roughly estimate the recombination rate using the recombination coefficient $\alpha_{\rm rec}$, which are smaller than $\lesssim 10^{-12}~{\rm cm^3~s^{-1}}$ at $10^4$ K for the listed elements \citep[Table 14.7 of][]{Draine11}.
We set the temperature of the stellar wind of $\approx 10^4$ K \citep[e.g., for the solar wind,][]{Wu&Judge79, Burlaga+02}, although the stellar wind electrons move supersonically with respect to NISPs.
Using $n_{\rm sw \odot} \approx 2.3 \times 10^{5}~{\rm cm^{-3}}$, we obtain the recombination rate $\alpha_{\rm rec} n_{\rm sw \odot} \lesssim 10^{-6}~{\rm s^{-1}}$ that is much lower than the photoionization rate $\beta_{{\rm ph}\odot, i} \gtrsim 10^{-3}~{\rm s^{-1}}$ (the red bars in the right panel of Figure \ref{fig:SolarSpectrum}).
Thus, for simplicity, we neglect the recombination process in Equation (\ref{eq:IonizationEquation}).


We obtain the analytic solution of Equation (\ref{eq:IonizationEquation}),
\begin{eqnarray}\label{eq:NeutralDensityAnalytic}
	\frac{n_i(r)}{n_{{\rm ISM}, i}}
	& = &
	\exp
	\left[
		- \frac{\sqrt{2} \beta_{{\rm ph} \star,i}}{\Omega_{{\rm K} \star}}
		\left(
			\sqrt{\frac{v^2_{\rm rel}}{v^2_{\rm esc \star}} + \frac{R_{\star}}{r}} - \frac{v_{\rm rel}}{v_{\rm esc \star}}
		\right)
	\right].
\end{eqnarray}
Although the neutral fraction $n_i(r) / n_{{\rm ISM}, i}$ increases with $v_{\rm rel}$, the accretion rate itself is fairly small $\dot{M}_{\rm BHL} \propto v^{-3}_{\rm rel}$ when $v_{\rm rel}$ is large.
For $v_{\rm rel} \ll v_{\rm esc \star}$, the fraction of NISPs accreting onto the stellar surface is approximated as
\begin{eqnarray}\label{eq:NeutralDensityRatio}
	\frac{n_{\star,i}}{n_{{\rm ISM},i}}
	\approx
	\exp
	\left(
		- \frac{\sqrt{2} \beta_{{\rm ph} \star,i}}{\Omega_{\rm K,\star}}
	\right).
\end{eqnarray}
The ratio $\beta_{{\rm ph}\star,i} / \Omega_{\rm K \star}$ is needed to be smaller than unity for accretion of the considerable fraction of NISPs.

\subsection{Results}\label{sec:SecondResults}


\subsubsection{Photoionization Rate}\label{sec:PhotoionizationRate}

%
\begin{figure}
\begin{center}
	\includegraphics[scale=0.8]{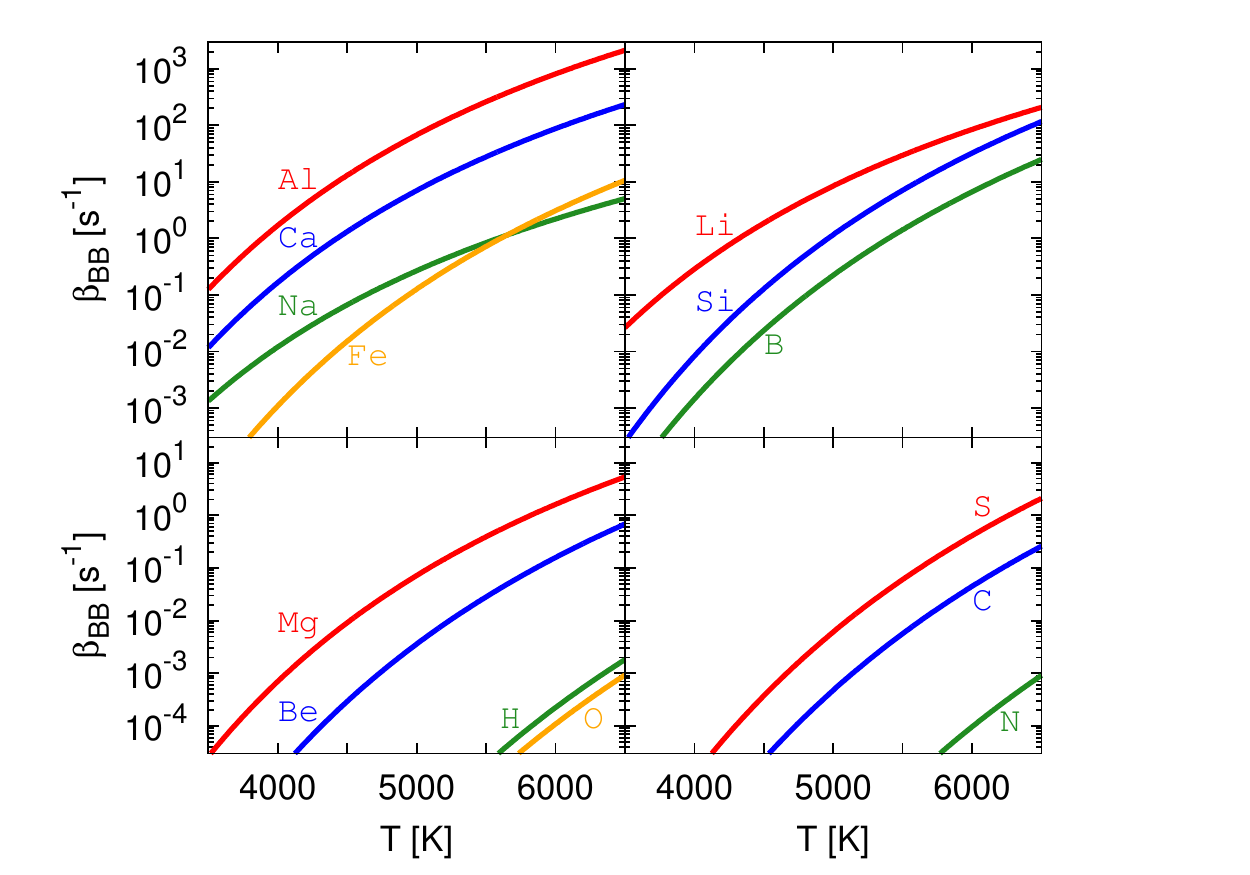}
\end{center}
\caption{
	The photoionization rate by the blackbody radiation is plotted for different elements.
	He, F, Ne, and Ar are not found in the plot because their $\beta_{{\rm BB},i}(T)$ are too small.
}
\label{fig:BBPhotoionizationRate}
\end{figure}
\begin{table}[!t]
\caption{
	The photoionization rate by the EUV component for different elements.
	$L_{{\rm EUV} \star,\lambda} = 10^{25}~{\rm erg~s^{-1}~nm^{-1}}$ and $R_{\star} = R_{\odot}$ are adopted.
}
\label{tbl:EUVPhotoionizationRate}
\begin{center}
\begin{tabular}{cl}
	Element $i$ & $\beta_{{\rm EUV} \odot,i}$ [s$^{-1}$]   \\

\hline

H       & $9.1 \times 10^{-4} $  \\ 
He      & $4.1 \times 10^{-4} $  \\ 
Li      & $3.0 \times 10^{-3} $  \\ 
Be      & $1.1 \times 10^{-3} $  \\ 
B       & $8.8 \times 10^{-3} $  \\ 
C       & $6.1 \times 10^{-3} $  \\ 
N       & $3.4 \times 10^{-3} $  \\ 
O       & $2.7 \times 10^{-3} $  \\ 
F       & $1.7 \times 10^{-3} $  \\ 
Ne      & $1.1 \times 10^{-3} $  \\ 
Na      & $1.3 \times 10^{-4} $  \\ 
Mg      & $5.7 \times 10^{-4} $  \\ 
Al      & $3.4 \times 10^{-2} $  \\ 
Si      & $2.5 \times 10^{-2} $  \\ 
S       & $1.8 \times 10^{-2} $  \\ 
Ar      & $6.6 \times 10^{-3} $  \\ 
Ca      & $4.0 \times 10^{-3} $  \\ 
Fe      & $3.0 \times 10^{-3} $  \\ 

\end{tabular}
\end{center}

\end{table}

The rates of photoionization by the blackbody and the EUV components are calculated separately.
We rewrite Equations (\ref{eq:PhotoionizationRateAtSurface}) and (\ref{eq:IntensityOfPopIII}) into
\begin{eqnarray}\label{eq:PhotoionizaionModel}
	\beta_{{\rm ph} \star,i}
	& = &
	\beta_{{\rm BB},i}(T) \nonumber \\
	& + &
	\beta_{{\rm EUV} \odot,i}
	\left( \frac{L_{{\rm EUV} \star,\lambda}}{10^{25}~{\rm erg~s^{-1}~nm^{-1}}} \right)
	\left( \frac{R_{\star}                  }{R_{\odot}                       } \right)^{-2}.
\end{eqnarray}
The rate by the blackbody $\beta_{{\rm BB},i}(T)$ is only a function of temperature $T$ and the calculated values for different elements are plotted in Figure \ref{fig:BBPhotoionizationRate}.
On the other hand, in order to determine the rate by the EUV component, we need to model both $L_{{\rm EUV} \star,\lambda}$ and $R_{\star}$.
The normalization $\beta_{{\rm EUV} \odot,i}$ in Equation (\ref{eq:PhotoionizaionModel}) is given in Table \ref{tbl:EUVPhotoionizationRate}.
According to Equation (\ref{eq:NeutralDensityRatio}), the plotted and tabulated $\beta_{{\rm ph \star},i}$ in Figure \ref{fig:BBPhotoionizationRate} and Table \ref{tbl:EUVPhotoionizationRate} should be compared with $\Omega_{\rm K \star} \sim$ a few $\times~10^{-3}~{\rm rad~s^{-1}}$ on the bottom panel of Figure \ref{fig:StellarModel}.

Some low ionization potential elements, lithium, aluminum, calcium, iron and so on, are predominantly ionized by the stellar blackbody radiation and then we do not need to take into account the EUV components.
For 0.7 $M_{\odot}$ PopIII stars ($T_{\rm eff} \sim$ 5700 K), most elements are primarily ionized by the Wien's tail of the blackbody radiation rather than the EUV component.
Even for 0.2 $M_{\odot}$ PopIII stars ($T_{\rm eff} \sim$ 4100 K), lithium, sodium, magnesium, aluminum, silicon, calcium, and iron are predominantly ionized by the blackbody radiation.

\subsubsection{Survival Probability}\label{sec:SurvivalProbility}

%
\begin{figure}
\begin{center}
	\includegraphics[scale=0.7]{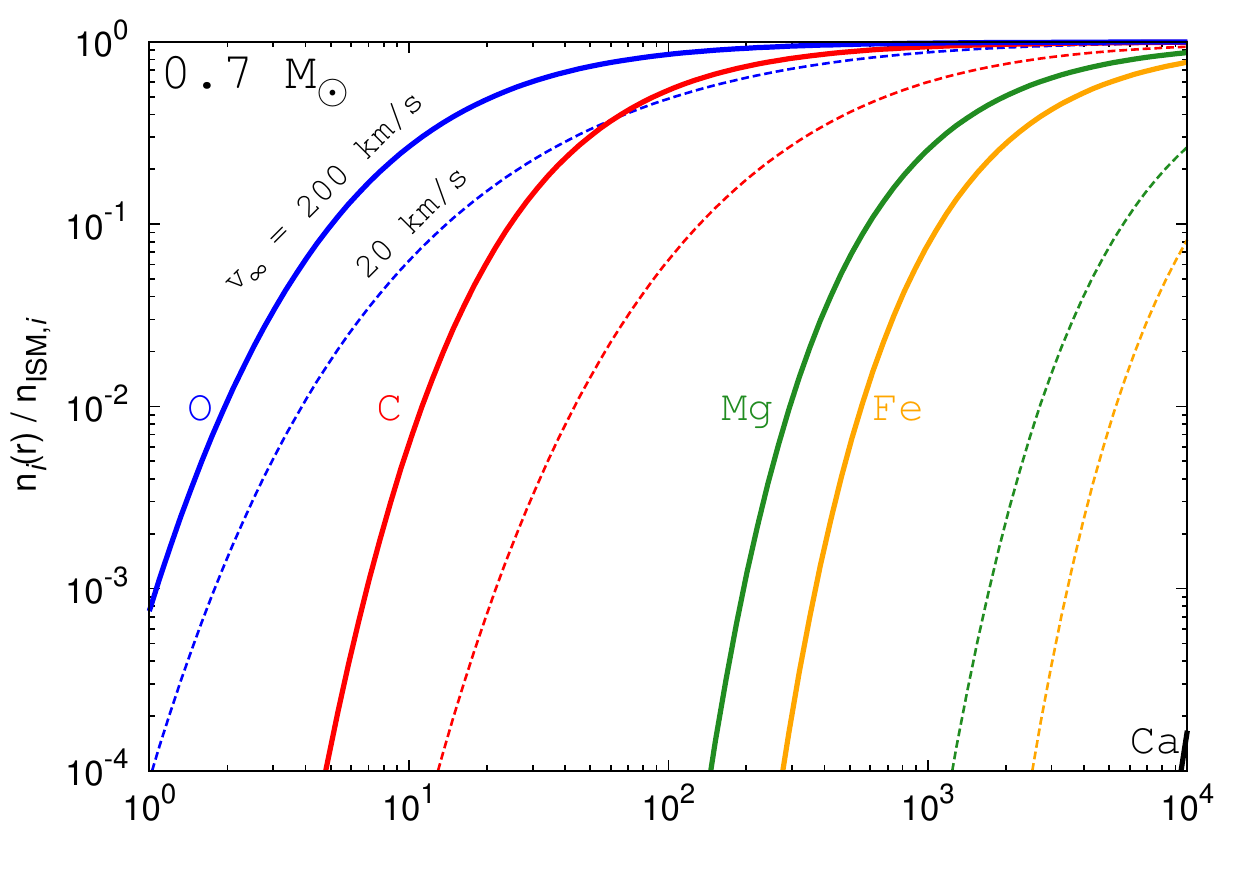} \\
	\includegraphics[scale=0.7]{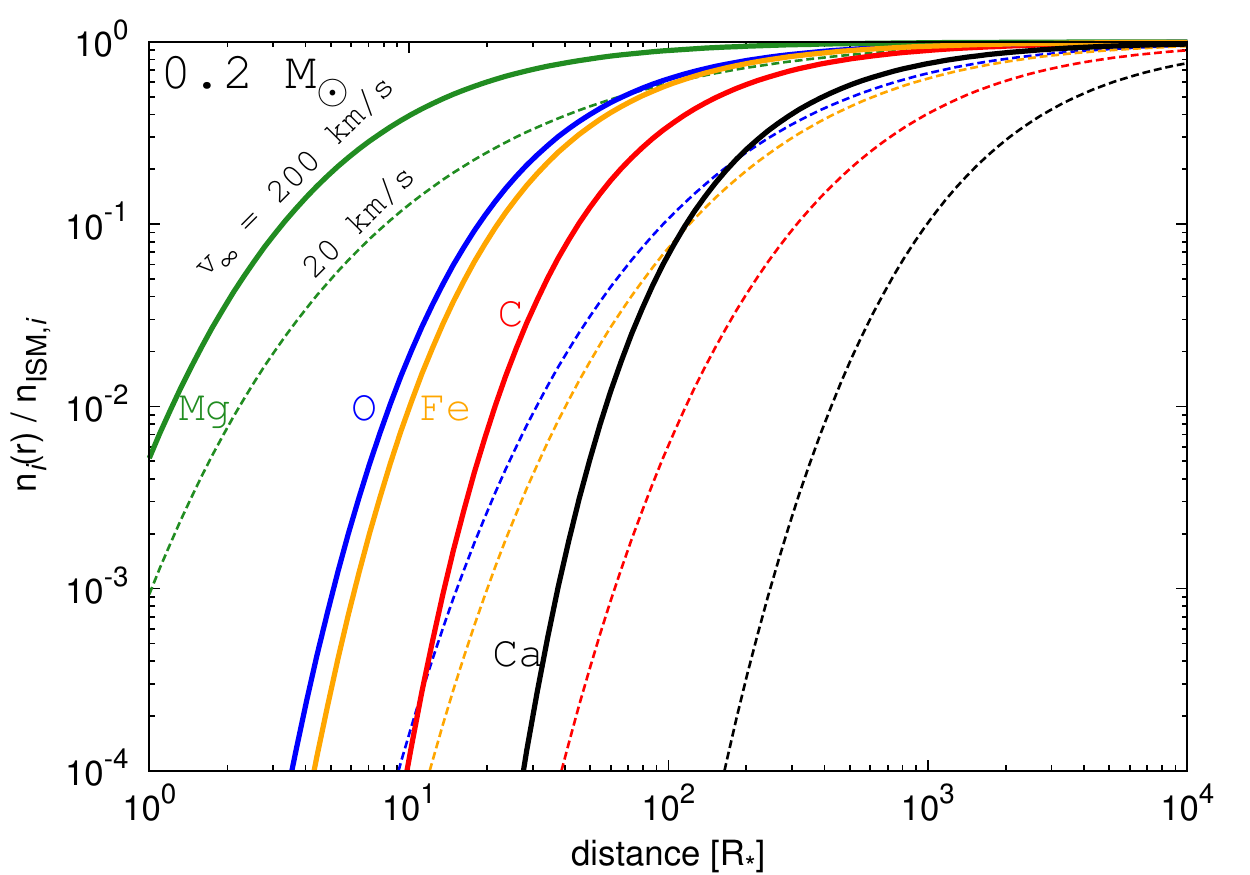} \\
\end{center}
\caption{
	The radial density profile of the neutral particles penetrating into the stellar magnetosphere (Equation (\ref{eq:NeutralDensityAnalytic})).
	Upper and lower panels are based on the model of a 0.7 $M_{\odot}$ and a 0.2 $M_{\odot}$ ZAMS PopIII star, respectively (Section \ref{sec:StellarModel}).
	The EUV component is $L_{{\rm EUV} \star,\lambda} = 10^{25}~{\rm erg~s^{-1}~nm^{-1}}$ for the both panels.
	The plots are for representative elements, where carbon (C:red), oxygen (O:blue), magnesium (Mg:green), iron (Fe:yellow), and calcium (Ca:black) are drawn.
	For solid lines, we adopt $v_{\rm rel} = 200~{\rm km~s^{-1}}$ considering the relative velocity of the gas and stars from \citet{Shen+16}.
	For dotted lines, $v_{\rm rel} = 20~{\rm km~s^{-1}}$ is about the velocity of the Sun against the local ISM.
}
\label{fig:DensityProfile}
\end{figure}
\begin{table*}[!t]
\caption{
	The logarithm of the survival probability at the stellar surface $\log(n_{\star,i}/n_{{\rm ISM},i})$ for different elements, where $L_{{\rm EUV} \star,\lambda} = 10^{25}~{\rm erg~s^{-1}~nm^{-1}}$.
}
\label{tbl:PhotoRatio}
\begin{center}
\begin{tabular}{cllllllll}
Element & our Sun         & 0.7$M_{\odot}$\footnotemark & 0.7 $M_{\odot}$       & 0.6 $M_{\odot}$       & 0.5 $M_{\odot}$       & 0.4 $M_{\odot}$       & 0.3 $M_{\odot}$       & 0.2 $M_{\odot}$       \\

\hline
$i$     & \multicolumn{8}{c}{$\log(n_{\star,i}/n_{{\rm ISM},i})$}    \\
\hline
H       & $-4.8               $ & $-1.6               $ & $-1.5               $ & $-1.7               $ & $-2.0               $ & $-2.5               $ & $-3.2               $ & $-4.6               $ \\
He      & $-3.9               $ & $-0.57              $ & $-0.64              $ & $-0.75              $ & $-0.90              $ & $-1.1               $ & $-1.4               $ & $-2.1               $ \\
Li      & $-8.4 \times 10^{ 3}$ & $-8.4 \times 10^{ 4}$ & $-2.3 \times 10^{ 4}$ & $-4.4 \times 10^{ 3}$ & $-1.2 \times 10^{ 3}$ & $-5.1 \times 10^{ 2}$ & $-2.7 \times 10^{ 2}$ & $-1.1 \times 10^{ 2}$ \\
Be      & $-3.2 \times 10^{ 1}$ & $-1.9 \times 10^{ 2}$ & $-3.0 \times 10^{ 1}$ & $-4.1               $ & $-2.8               $ & $-3.2               $ & $-4.0               $ & $-5.7               $ \\
B       & $-2.6 \times 10^{ 2}$ & $-7.6 \times 10^{ 3}$ & $-1.3 \times 10^{ 3}$ & $-1.4 \times 10^{ 2}$ & $-3.8 \times 10^{ 1}$ & $-3.0 \times 10^{ 1}$ & $-3.3 \times 10^{ 1}$ & $-4.5 \times 10^{ 1}$ \\
C       & $-4.0 \times 10^{ 1}$ & $-6.8 \times 10^{ 1}$ & $-1.6 \times 10^{ 1}$ & $-1.1 \times 10^{ 1}$ & $-1.3 \times 10^{ 1}$ & $-1.7 \times 10^{ 1}$ & $-2.1 \times 10^{ 1}$ & $-3.0 \times 10^{ 1}$ \\
N       & $-1.6 \times 10^{ 1}$ & $-4.9               $ & $-5.4               $ & $-6.2               $ & $-7.5               $ & $-9.4               $ & $-1.2 \times 10^{ 1}$ & $-1.7 \times 10^{ 1}$ \\
O       & $-1.6 \times 10^{ 1}$ & $-3.9               $ & $-4.2               $ & $-4.8               $ & $-5.8               $ & $-7.3               $ & $-9.3               $ & $-1.3 \times 10^{ 1}$ \\
F       & $-1.4 \times 10^{ 1}$ & $-2.4               $ & $-2.7               $ & $-3.1               $ & $-3.8               $ & $-4.7               $ & $-6.0               $ & $-8.6               $ \\
Ne      & $-1.3 \times 10^{ 1}$ & $-1.5               $ & $-1.6               $ & $-1.9               $ & $-2.3               $ & $-2.9               $ & $-3.7               $ & $-5.3               $ \\
Na      & $-2.4 \times 10^{ 2}$ & $-2.1 \times 10^{ 3}$ & $-6.2 \times 10^{ 2}$ & $-1.4 \times 10^{ 2}$ & $-4.0 \times 10^{ 1}$ & $-1.8 \times 10^{ 1}$ & $-1.0 \times 10^{ 1}$ & $-4.3               $ \\
Mg      & $-2.3 \times 10^{ 1}$ & $-1.8 \times 10^{ 3}$ & $-3.5 \times 10^{ 2}$ & $-4.0 \times 10^{ 1}$ & $-8.2               $ & $-3.9               $ & $-3.0               $ & $-3.1               $ \\
Al      & $-6.2 \times 10^{ 4}$ & $-8.2 \times 10^{ 5}$ & $-2.1 \times 10^{ 5}$ & $-3.5 \times 10^{ 4}$ & $-8.6 \times 10^{ 3}$ & $-3.6 \times 10^{ 3}$ & $-1.8 \times 10^{ 3}$ & $-7.4 \times 10^{ 2}$ \\
Si      & $-7.8 \times 10^{ 2}$ & $-3.6 \times 10^{ 4}$ & $-6.6 \times 10^{ 3}$ & $-6.9 \times 10^{ 2}$ & $-1.6 \times 10^{ 2}$ & $-1.0 \times 10^{ 2}$ & $-1.0 \times 10^{ 2}$ & $-1.3 \times 10^{ 2}$ \\
S       & $-1.1 \times 10^{ 2}$ & $-5.5 \times 10^{ 2}$ & $-9.4 \times 10^{ 1}$ & $-3.7 \times 10^{ 1}$ & $-4.0 \times 10^{ 1}$ & $-5.0 \times 10^{ 1}$ & $-6.4 \times 10^{ 1}$ & $-9.1 \times 10^{ 1}$ \\
Ar      & $-2.2 \times 10^{ 1}$ & $-9.2               $ & $-1.0 \times 10^{ 1}$ & $-1.2 \times 10^{ 1}$ & $-1.4 \times 10^{ 1}$ & $-1.8 \times 10^{ 1}$ & $-2.3 \times 10^{ 1}$ & $-3.3 \times 10^{ 1}$ \\
Ca      & $-4.3 \times 10^{ 3}$ & $-8.9 \times 10^{ 4}$ & $-2.2 \times 10^{ 4}$ & $-3.7 \times 10^{ 3}$ & $-8.8 \times 10^{ 2}$ & $-3.6 \times 10^{ 2}$ & $-1.9 \times 10^{ 2}$ & $-7.6 \times 10^{ 1}$ \\
Fe      & $-7.5 \times 10^{ 1}$ & $-3.4 \times 10^{ 3}$ & $-6.5 \times 10^{ 2}$ & $-7.5 \times 10^{ 1}$ & $-1.8 \times 10^{ 1}$ & $-1.2 \times 10^{ 1}$ & $-1.2 \times 10^{ 1}$ & $-1.5 \times 10^{ 1}$ \\

\end{tabular}
\end{center}

\footnotetext[1]{
	Calculated for a 0.7 $M_{\odot}$ star at an age of 13.8 Gyr.
}
\end{table*}
%

The left-hand side of Equation (\ref{eq:NeutralDensityAnalytic}) corresponds to the `survival probability', i.e., the neutral fraction of NISPs \citep[e.g.,][]{Bzowski+13a}.
Figure \ref{fig:DensityProfile} shows the radial distribution of carbon, oxygen, magnesium, iron, and calcium atoms (Equation (\ref{eq:NeutralDensityAnalytic})) for a 0.7 $M_{\odot}$ ZAMS PopIII star (upper panel) and for a 0.2 $M_{\odot}$ ZAMS PopIII star (bottom panel).
For the case of a 0.7 $M_{\odot}$ PopIII star, oxygen (blue) has the largest and calcium has the smallest survival probabilities among the five elements.
However, magnesium is the largest for the case of a 0.2 $M_{\odot}$ PopIII star because oxygen and carbon are primarily ionized by the EUV component $\beta_{{\rm EUV} \star,i}$ in this case.
The abundance ratio of the accreting medium will be significantly different from the surrounding ISM because the survival probabilities at the stellar surface strongly depend on elements and also stellar masses.

In Table \ref{tbl:PhotoRatio}, we show a table of $\log (n_{\star,i} / n_{{\rm ISM},i})$ calculated with Equation (\ref{eq:NeutralDensityRatio}).
For $v_{\rm rel} = 20~{\rm km~s^{-1}}$, the tabulated values almost correspond to the survival probabilities at $r = R_{\star}$ in Figure \ref{fig:DensityProfile} (dotted blue line on the upper panel and the dotted green line on the bottom panel).
For $v_{\rm rel} = 200~{\rm km~s^{-1}}$, the survival probabilities (solid lines in Figure \ref{fig:DensityProfile}) become a bit higher than the case of $v_{\rm rel} = 20~{\rm km~s^{-1}}$ (dotted lines in Figure \ref{fig:DensityProfile}).
For $v_{\rm esc} = 680~{\rm km~s^{-1}}$ (0.7 $M_{\odot}$), the parentheses of Equation (\ref{eq:NeutralDensityAnalytic}) is $\sqrt{v^2_{\rm rel}/v^2_{\rm esc \star} + 1} - v_{\rm rel}/v_{\rm esc \star} \approx$ 0.97 ($v_{\rm rel} = 20~{\rm km~s^{-1}}$) and 0.75 ($v_{\rm rel} = 200~{\rm km~s^{-1}}$), respectively.
Only hydrogen, nitrogen, oxygen, fluorine and noble gases have the survival probability of $> 10^{-10}$ at the surface of 0.7 $M_{\odot}$ PopIII stars.

\citet{Shen+16} argued that BHL accretion of metal-enriched gases onto low-mass PopIII stars can enrich them $-10 \lesssim {\rm [Fe/H]}_{\rm Shen} \lesssim -2$ with the median value of $\sim -5$.
As already discussed in Section \ref{sec:FirstResults}, the stellar magnetosphere is safely formed in entire their evolution so that only survived iron of NISPs would accrete onto the stars.
Although we studied only the one-dimensional trajectory of a NISP, we may crudely estimate the accreting iron abundance as ${\rm [Fe/H]}_{\rm Shen} + \log(n_{\star,{\rm Fe}} / n_{{\rm ISM,Fe}})$.
From Table \ref{tbl:PhotoRatio}, we find $\log(n_{\star,{\rm Fe}} / n_{{\rm ISM,Fe}}) < -12$ and then the iron enrichment by BHL accretion becomes ${\rm [Fe/H]} < -14$, which is far less than the observed values \citep[e.g.,][]{Christlieb+04, Aoki+06, Caffau+11, Keller+14, Frebel+15}.

\subsection{Discussion}\label{sec:SecondDiscussion}

\subsubsection{Uncertainties of EUV component}\label{sec:EUVComponent}

The resultant survival probabilities depend on the assumption on the specific intensity, especially the EUV component of low-mass PopIII stars.
However, since lithium, aluminum, calcium, iron are primarily ionized by the blackbody component, these elements hardly accrete onto low-mass PopIII stars even for a smaller EUV component than $L_{{\rm EUV} \star,\lambda} = 10^{25}~{\rm erg~s^{-1}~nm^{-1}}$.
On the other hand, for some elements, such as oxygen and carbon in Figure \ref{fig:DensityProfile}, the weak EUV component may allow them to attain the stellar surface.

\subsubsection{Charge Exchange}\label{sec:ChargeExchange}

%
\begin{figure}
\begin{center}
	\includegraphics[scale=0.7]{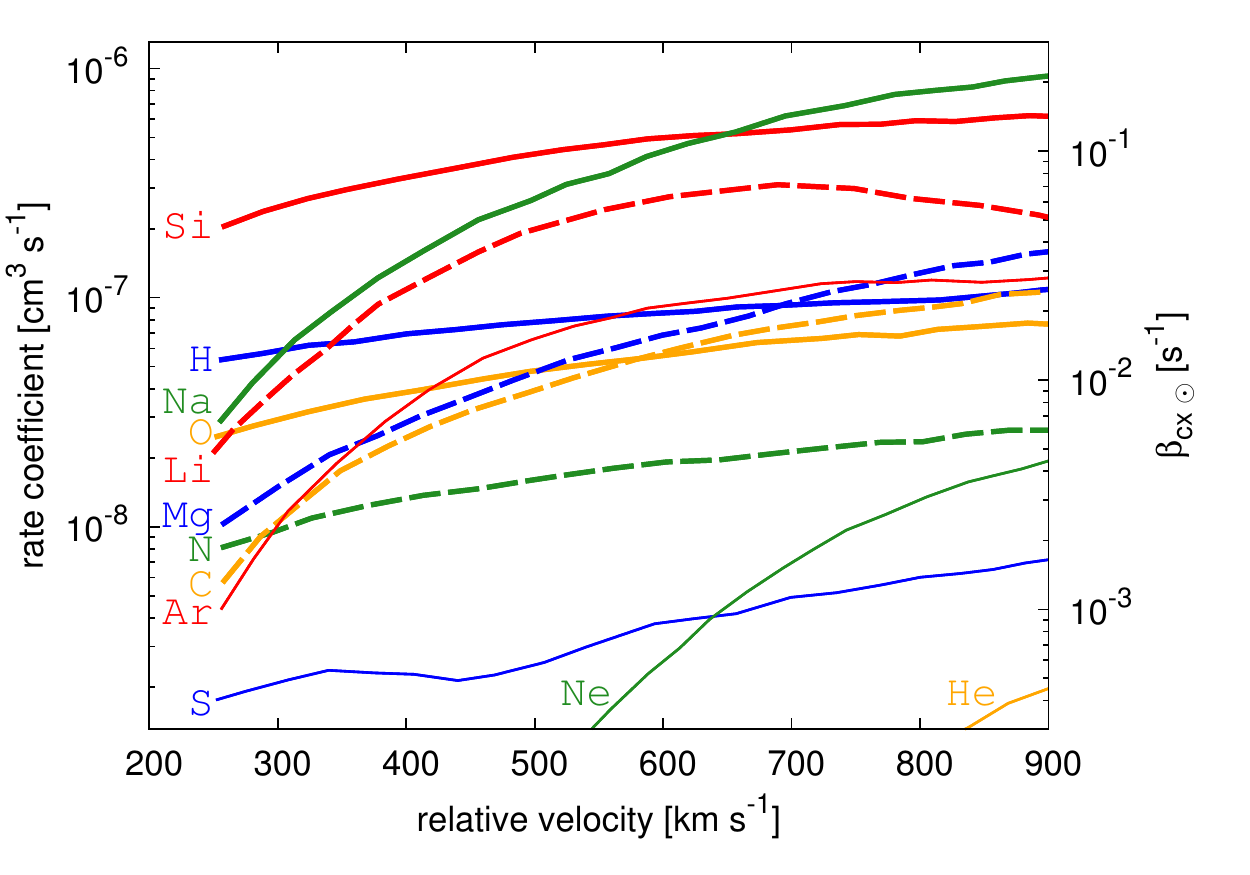}
\end{center}
\caption{
	The rate coefficients of charge exchange processes with neutral atom and proton ($\rm X + H^+ \rightarrow X^+ + H$).
	Data for Li is taken from \citet{Barnett90} and for the others are from \citet{Cummings+02}.
	We are interested in the relative velocity of $\sim v_{\rm sw \star} = 400~{\rm km~s^{-1}}$.
	The right scale is the ionization rate by the charge exchange process at the solar surface $\beta_{{\rm cx}\odot, i} \equiv \sigma_{{\rm cx},i} v_{\rm rel} n_{\rm sw  \odot}$.
}
\label{fig:ChargeExchange}
\end{figure}
%



NISPs are also suffered from ionization by charge exchange with stellar wind ions (mainly proton) and from electron impact ionization by stellar wind electrons \citep[c.f.,][]{Zank99}.
These processes increase the total ionization rate and make the metal enrichment more difficult.
In the case of the heliosphere, the charge exchange is the dominant ionization process only for the hydrogen, while the heavy elements are primarily photoionized \citep[c.f.,][]{Cummings+02, Bzowski+13a}.
However, if the EUV component of the stellar radiation is significantly weak, these processes will play an important role for some elements.

Figure \ref{fig:ChargeExchange} is a plot of the rate coefficients of the charge exchange between a neutral atom X and a proton ($\rm X + H^+ \rightarrow X^+ + H$).
The cross section $\sigma_{{\rm cx}, i}$ is a function of the relative velocity of the particles $w$ ($w = v_{\rm rel} + v_{\rm sw \star}$ for head-on collision of NISPs and stellar wind plasma), where the rate coefficient $\equiv \sigma_{{\rm cx}, i} w$ is found on the left scale of the plot.
In order to estimate the ionization rate by the change exchange, we need the density and velocity profiles of the stellar wind of low-mass PopIII stars.
The right scale of Figure \ref{fig:ChargeExchange} is $\beta_{{\rm cx}\odot,i} \equiv \sigma_{{\rm cx}, i} w n_{\rm sw \odot}$ for reference, and the values are comparable to $\beta_{{\rm ph}\odot, i}$ on the right panel of Figure \ref{fig:SolarSpectrum}.
Note that the charge exchange cross section for the neutral iron is not studied in the relative velocity of our interest \citep[c.f.,][]{Patton+94}.
\section{General Discussion}\label{sec:Discussion}



In this paper, we only study metal accretion in gas phase.
Some of heavy elements are contained in dust and this could accrete, if angular momentum of the dust is successfully extracted.
Otherwise, the accretion cross section is just a geometrical cross section.
In addition, \citet{Johnson15} discussed that the radiation pressure force prevents the accretion of the dust whose grain size ranges $\sim 10^{-2} - 1~{\rm \mu m}$.
The dust distribution inside the heliosphere has also been studied and some studies suggest that charging of dust particles also plays a role to prevent penetrating into the inner magnetosphere \citep[][]{Bertaux&Blamont76, Levy&Jokipii76, Mann10}.
More detailed studies for the dust accretion is left for future studies.

\citet{Talbot&Newman77} adopted the condition $R_{\rm TS} > \xi_{\rm BHL} (\gg R_{\star})$ rather than $R_{\rm TS} > R_{\star}$ to obtain the critical density $n_{\rm crit}$ (Equation (\ref{eq:CriticalDensity})).
They argued that there is no stable radius where the ram pressures of the wind and the accretion flow balances between $R_{\star}$ and $\xi_{\rm BHL} = R_{\star} (v_{\rm esc} / v_{\rm rel})^2$.
However, they assumed the spherical free-fall trajectory of the accretion flow $n v^2 \propto r^{-5/2}$ at $r < \xi_{\rm BHL}$, and did not consider the axisymmetric trajectory of Equation (\ref{eq:HoyleLyttletonTrajectory}).
Both of the spherical and axisymmetric trajectories will be modified by including the interactions of neutrals in the accretion flow with the stellar wind.
For example, \citet{Wallis71} and \citet{Holzer72} discussed a shock-free magnetosphere because of gradual deceleration and heating of the wind by NISPs.

Accretion onto objects which have a wind mass loss requires further discussion to determine $n_{\rm crit}$ even without neutrals.
BHL accretion takes place even from the downwind direction \citep[e.g.,][]{Shima+85, Ruffert96}.
The accretion flow might be compressed by forming shock behind the star and by radiative cooling.
The cooled dense gas may accrete onto the star from the magnetotail region and the density, and thus the ram pressure, of accreting flow might effectively exceed $n_{\rm crit}$.
However, the flow pattern of BHL accretion itself should be modified by the existence of the stellar wind, for example, the accretion shock region behind the star might be further away from the star than the case without the stellar wind.
We need further studies for BHL accretion of the mostly neutral ISM including interactions with the stellar wind.

The stellar wind model adopted in the present paper is based on the studies of nearby low-mass MS stars.
Even for low-mass MS stars, more observational constraints are required to develop a model of their wind, although there are some comprehensive numerical studies \citep[e.g.,][]{Pinto+11, Cranmer&Saar11, Cohen&Drake14, Johnstone+15b, Johnstone+15a, Johnstone17, Reville+15}.
In addition to the magnetic and rotation activities of stars, the radiative cooling processes of the stellar atmosphere could be different between the observed low-mass MS stars and low-mass PopIII stars, and thus the wind properties would be significantly affected \citep[e.g.,][]{Suzuki&Inutsuka05, Suzuki&Inutsuka06, Matsumoto&Suzuki14}.
Further studies of the stellar wind from low-mass PopIII stars itself are also required.

As discussed in Section \ref{sec:MagnetosphereFormation}, low-mass PopIII stars maintain the magnetosphere around them in most of their life.
In the presence of the magnetosphere, photoionization of NISPs reduces metal enrichment by accretion to undetectable level [Fe/H] $< -14$ \citep[][]{Frebel+09}, while the abundance of nitrogen, oxygen, fluorine and noble gases can be higher than the other elements.
For accretion from an overdense region, \citet{Johnson&Khochfar11} gives one-tenth of low-mass PopIII stars can be metal-enriched to the observed lowest [Fe/H] $\sim -5$.
In other words, if the three stars of [Fe/H] $< -5$ were such lucky low-mass PopIII stars \citep[][]{Christlieb+04, Aoki+06, Frebel+15}, we expect that at least thirty metal free stars should have been observed till now but we have not.
Combining these results, it is difficult to interpret the observed metal poor stars as metal-enriched low-mass PopIII stars.
Thus, the observations of metal poor stars indicate the top-heavy IMF for PopIII \citep[][]{Hartwig+15, Ishiyama+16}.


\section{Conclusions}\label{sec:Conclusions}

The effects of the stellar wind on the metal accretion onto low-mass PopIII stars are investigated.
We model the stellar wind and radiation of low-mass PopIII stars based on the observations of the heliosphere and some low-mass stellar systems close to the Sun.
The distribution of the local interstellar neutral atoms in the heliosphere has been extensively studied and we apply the similar but fairly simplified discussion to low-mass PopIII stars.
We find that the metal enrichment by BHL accretion of the ionized ISM against their stellar wind rarely happens.
Although the neutrals can penetrate into the magnetosphere, most neutral atoms are ionized before reaching the stellar surface and will be picked up by the stellar wind which blows the interstellar heavy elements away to the outer magnetosphere.

Formation of magnetosphere plays a crucial role in the present results.
We find that stars moving dense region of $\gtrsim 10^4~{\rm cm^{-3}}$ would be ceased to form the stellar magnetosphere and be metal enriched by accretion.
However, such dense region occupies a small fraction of the volume of the Galaxy.
It is hardly expected for low-mass PopIII stars to go through the dense cloud and the magnetosphere is formed around low-mass PopIII stars in most of their life.


We also consider metal enrichment by neutrals in the ISM, because they do not interact electromagnetically with the stellar wind plasma.
However, once the stellar magnetosphere is formed, the neutral interstellar heavy elements in gas phase hardly accrete onto low-mass PopIII stars because of photoionization.
Although some studies predict that the chemical enrichment by BHL accretion accomplishes ${\rm [Fe/H]} \sim -2$ in extreme cases, the accretion rate of the interstellar iron is reduced by $< 10^{-12}$ times, i.e., ${\rm [Fe/H]} \sim -14$ even in the extreme case.

The chemical enrichment by BHL accretion is unlikely to explain the observations of the metal poor stars, i.e., they are born with the present element abundances.
In other words, low-mass PopIII stars remain pristine to be found as metal free stars.

\section*{Acknowledgments}

S. J. T. would like to thank Y. Ohira, T. Terasawa, T. Suzuki, D. Kinoshita, K. Takaharashi, T. Hartwig, K. Omukai, and T. Hosokawa for useful and helpful discussions.
The authors would also like to thank the anonymous referee for useful and helpful comments.
This work is supported by Grants-in-Aid for Scientific Research Nos. 17K18270 (ST), 15H05440 (NT), 17H02869 (HS) and 17H01101 (HS).
G. C. is supported by JSPS Research Fellowships for Young Scientists.

\bibliography{draft}

\end{document}